\documentclass[twocolumn,superscriptaddress,showpacs,pra]{revtex4}
\usepackage{epsfig}
\usepackage{delarray}
\usepackage{amsmath, amssymb}
\usepackage{bm}
\usepackage{graphicx}
\usepackage{placeins}
\usepackage{color}

\begin{document}
\title{Rogue waves of the Kundu-Eckhaus equation in a chaotic wave field}

\author{Cihan Bayindir}
\email{cihan.bayindir@isikun.edu.tr}
\affiliation{Department of Civil Engineering, Isik University, Istanbul, Turkey}

\begin{abstract}
In this paper we study the properties of the chaotic wave fields generated in the frame of the Kundu-Eckhaus equation (KEE). Modulation instability results in a chaotic wave field which exhibits small-scale filaments with a free propagation constant, k. The average velocity of the filaments is approximately given by the average group velocity calculated from the dispersion relation for the plane-wave solution however direction of propagation is controlled by the $\beta$ parameter, the constant in front of the Raman-effect term. We have also calculated the probabilities of the rogue wave occurrence for various values of propagation constant k and showed that the probability of rogue wave occurrence depends on k. Additionally, we have showed that the probability of rogue wave occurrence significantly depends on the quintic and the Raman-effect nonlinear terms of the KEE. Statistical comparisons between the KEE and the cubic nonlinear Schr\"{o}dinger equation have also been presented.

\pacs{05.45.-a, 05.45.-Yv, 02.60.Cb}
\end{abstract}
\maketitle

\section{\label{sec:level1} Introduction}
Rogue (freak) wave studies have become extensive in recent years \cite{Akhmediev2011, bayindir2016}. These studies has a great importance for the safety of the marine travel and offshore operations as it is important to avoid rogue waves in the open ocean. Rogue wave studies are also crucial for many other fields such as optics, dynamics of super fluids and finance just to name a few \cite{Wang}. Researchers working in various areas want to understand the physics behind the rogue wave phenomenon.

The research has naturally started with the investigation of one of the simplest nonlinear mathematical models, which is the nonlinear Schr\"{o}dinger equation (NLSE) \cite{Akhmediev2014}. Discovery of the unexpected rogue wave solutions, even for this well-known equation resulted in seminal studies of rogue wave dynamics, such as \cite{Akhmediev2009b}. However, the NLSE has limitations due to the assumptions and approximations used in its derivation. Deriving or extending the solutions to more general dynamic equations will be the next step in rogue wave research.

The Kundu-Eckhaus equation (KEE) in one of the integrable extensions of the NLSE. It contains extension terms to the NLSE, namely the quintic and Raman-effect nonlinear terms \cite{Wang}. One of the versions of the KEE can be written in the form of
\begin{equation}
i\psi_t + \psi_{xx} + 2 \left|\psi \right|^2 \psi + \beta^2 \left|\psi \right|^4 \psi - 2 \beta i \left( \left|\psi \right|^2 \right)_x \psi =0
\label{eq01}
\end{equation}
where $x,t$ is the spatial and temporal variables, $i$ denotes the imaginary number and $\psi$ is complex amplitude \cite{Wang}. $\beta$ is a real constant and $\beta^2$ is the coefficient of the quintic nonlinear term which model the effects of higher order nonlinearity. The last term represents the Raman-effect which account for the self-frequency shift of the pulses \cite{Wang}. KEE equation can adequately model the propagation of ultrashort pulses in nonlinear and quantum optics. In mechanics, KEE is capable of examining the stability of Stokes waves in weakly nonlinear dispersive media.

\section{\label{sec:level1}Plane-Wave Solution}
Analytical rogue wave solutions always retain on a plane-wave background since the plane-waves serve as a energy source for the rogue waves \cite{Akhmediev2014}. Thus it is important to study the plane-wave solution of the KEE. The KEE given in Eq.\ref{eq01} admits a plane wave solution in the form of
\begin{equation}
\psi_o= \exp{\left( i\left[kx-\omega t \right]  \right)}=\exp{\left( i\left[kx-(k^2-\beta^2 -2)t  \right]  \right)}
\label{eq04}
\end{equation}

\begin{figure}[htb!]
\begin{center}
   \includegraphics[width=3.4in]{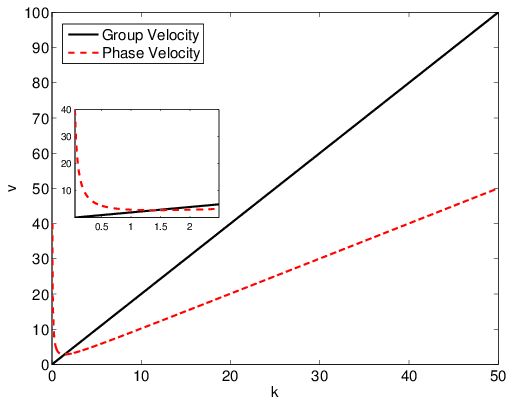}
  \end{center}
\caption{\small (Color online) Group ({\color{black} ---}) and phase ({\color{red} - -}) velocities of the plane wave for $\beta=0$.}
  \label{fig1}
\end{figure}

\noindent This solution has two free parameters $\beta$ and $k$. Using this plane-wave solution it can be easily deduced that the phase velocity of the plane-wave is given by
\begin{equation}
v_{ph}= \frac{\omega}{k}= k-(\beta^2 -2)/k  
\label{eq05}
\end{equation}
On the other hand the group velocity becomes
\begin{equation}
v_{gr}= \frac{\partial \omega}{\partial k}= 2k
\label{eq06}
\end{equation}
It is easy to check that if $ k \geq \sqrt{2-\beta^2}$ then $ v_{gr} \geq v_{ph}$; else $ v_{gr} < v_{ph}$. Dependence of the velocities on the wavenumber, $k$, is shown in Fig.\ref{fig1} for $\beta=0$. In order to avoid evanescent modes and analyze the propagation dynamics of waves we select real $k$ values. The velocities of ultrashort pulses are higher than those of long pulses, as the Fig.\ref{fig1} confirms. This property of the KEE frame is important to study the propagation dynamics of ultrashort pulses in optics.

\section{\label{sec:level1}Rogue Waves Solutions}

Setting $\beta=0$, the KEE reduces to the cubic NLSE for which the rogue wave solutions become obvious \cite{Akhmediev2009b}. For the cubic NLSE, the first order rational rogue wave solution is given by Peregrine in 1979 \cite{Peregrine}. Similar first order rational rogue wave solution for the KEE is recently presented in \cite{Zhao2013}. Second and the higher order rational solutions of the KEE and a hierarchy of obtaining those rational solutions based on Darboux transformations are given in \cite{Wang}. They are basically skewed rogue waves obtained by the gauge transforming the rogue wave solutions of the NLSE. For the sake of brevity we are not repeating their explicit formula here. However in order to present an illustration of the quintic and Raman-effect terms we present contour plots of the first order rational soliton solutions in the Fig. 2 for various $\beta$ values.

In Fig. 2, the contour plot of the first order rogue wave solution of the KEE is presented for various $\beta$ values. We can see that quintic and Raman-effect nonlinear terms produce an important skew angle relative to the ridge of the rogue waves. As also described in \cite{Wang}, the sign of the $\beta$ parameter determines the sign of the skew angle relative to the ridge of the rogue wave. If $\beta = 0$ then there is no skew angle and the rogue wave solution of the KEE is no different than the Peregrine soliton solution of the NLSE. If $\beta > 0$ the skew angle is in the counter clockwise direction. If $\beta < 0$ it is in the clockwise direction \cite{Wang}. Additionally as the magnitude of the $\beta$ gets larger, the skew angle gets larger as well \cite{Wang}. Illustration of this behavior for various values of $\beta$ helps us analyzing the behavior of the chaotic wave field, as described in the next section.

\begin{figure}[ht!]
\begin{center}
   \includegraphics[width=2.6in]{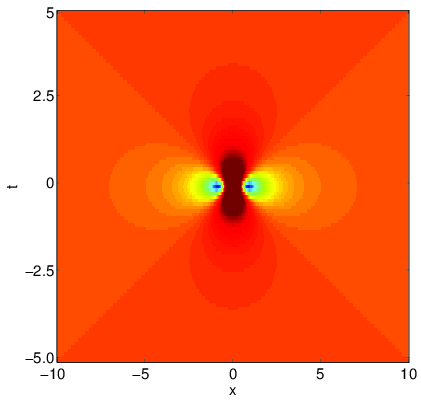}
  \end{center}
\end{figure}
\vspace{-1em}
\begin{figure}[h!]
\begin{center}
   \includegraphics[width=2.6in]{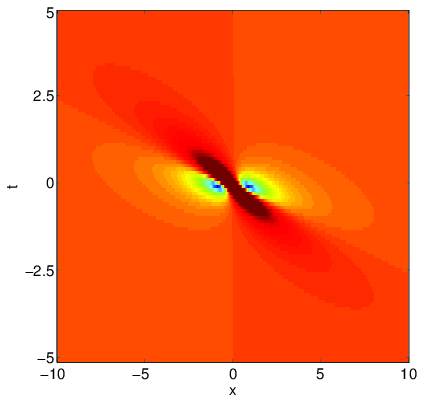}
  \end{center}
\end{figure}
\vspace{-1em}
\begin{figure}[h!]
\begin{center}
   \includegraphics[width=2.6in]{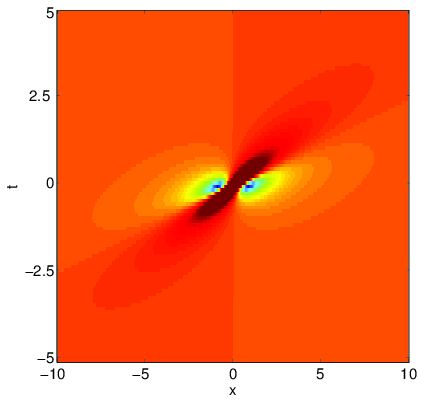}
  \end{center}
\caption{Contour plots of the first order rogue wave solution of the KEE a) for $\beta=0$, b) for $\beta=1$, c) for  $\beta=-1$.}
\end{figure}

\section{\label{sec:level1}Chaotic Wave Fields}

Although the processes governed by the KEE are very complicated, they are still governed by a partial differential equation. Therefore they can be predicted once an initial condition is specified. Thus compared to the completely unpredictable true stochastic processes, the processes described in the frame of KEE in this study can be named as 'chaotic' \cite{Akhmediev2014}.

In order to analyze chaotic wave-fields in the frame of the KEE we use a numerical framework. We start the rogue wave simulations using a constant amplitude wave with an additive small chaotic perturbation. Such a state is unstable and it evolves into a full-scale chaotic wave field similar with the numerical simulations described in \cite{Akhmediev2009b, Akhmediev2009a, bayindir2016}. The chaotic wave field modeled by the KEE with this starter evolves into a wave field which exhibits many amplitude peaks, with some of them becoming rogue waves. In order to model such a chaotic wave field we use the initial condition
\begin{equation}
\psi (x,t=0)=\psi_0 (x,0)+\mu a(x)
\label{eq08}
\end{equation}
where $\psi_0$ initial plane wave solution given by Eq. (\ref{eq04}), $a(x)$ is a uniformly distributed random complex function with real and imaginary parts have random values in the interval of $[-1,1]$. The actual water surface fluctuation for this initial condition would be given by $ \left|\psi\right| \exp{[i\omega t]}$ where $\omega$ is some carrier wave frequency.  Following \cite{bayindir2016}, a value of $\mu=0.2$ is selected. It is possible to add perturbations with a characteristic length scale $L_{pert}$, by multiplying the second term of the Eq. (\ref{eq08}) by a factor of $\exp(i 2 \pi / L_{pert}x)$. Or it is also add perturbations with different wavelength scales using Fourier analysis. However in the present study for illustrative purposes we do not specify such as scale. 

For the numerical solution of the KEE we propose and implement a split-step Fourier method (SSFM) as described below. In SSFM schemes, the spatial derivatives are evaluated using spectral techniques. Some of the applications of the spectral techniques can be seen in \cite{bayindir2009, Karjadi2010, Karjadi2012} and more detailed discussions can be seen in \cite{trefethen}. In spectral techniques the spatial derivatives are calculated by making use of the orthogonal transforms. The most popular choice for the periodic domains is the Fourier transform \cite{bayindir2016}. The time integration of the governing equation is performed using schemes such as Adams-Bashforth and Runge-Kutta etc. \cite{trefethen, demiray}. However for SSFM, exponential time stepping is used for time integration. 

SSFM relies on the idea of splitting the equation into two parts, the nonlinear and the linear part \cite{bayindir2015d, 
bayindir2015arxivchbloc, bayindir2015arxivcssfm, bay2015b}. For the KEE, the advance in time due to nonlinear part can be written as
\begin{equation}
i\psi_t= -(2\left| \psi \right|^2 +  \beta^2 \left| \psi \right|^4-2i\beta [\left| \psi \right|^2]_x    )\psi
\label{eq09}
\end{equation}
which can be exactly solved as
\begin{equation}
\tilde{\psi}(x,t_0+\Delta t)=e^{i(2\left| \psi_0 \right|^2 +  \beta^2 \left| \psi_0 \right|^4-2i\beta [\left| \psi_0 \right|^2]_x )\Delta t}\ \psi_0   
\label{eq10}
\end{equation}
where $\psi_0=\psi(x,t_0)$ is the initial condition, $\Delta t$ is the time step. It is possible to evaluate the spatial derivatives using the Fourier series so that we can write
\begin{equation}
\tilde{\psi}(x,t_0+\Delta t)=e^{i(2\left| \psi_0 \right|^2 +  \beta^2 \left| \psi_0 \right|^4-2i\beta F^{-1}[ikF[\left| \psi_0 \right|^2]] )\Delta t}\ \psi_0   
\label{eq10}
\end{equation}
where $k$ is the Fourier transform parameter and $F$ and $F^{-1}$ denote the forward and inverse Fourier transforms \cite{bayindir2016}. The linear part of the KEE can be written as
\begin{equation}
i\psi_t=-\psi_{xx}
\label{eq11}
\end{equation}
Using the Fourier series it is possible to write that
 \begin{equation}
\psi(x,t_0+\Delta t)=F^{-1} \left[e^{-ik^2\Delta t}F[\tilde{\psi}(x,t_0+\Delta t) ] \right]
\label{eq12}
\end{equation}
where $k$ is the Fourier transform parameter. Therefore combining Eq. (\ref{eq10}) and Eq. (\ref{eq12}), the complete form of the SSFM can be written as
 \begin{equation}
\begin{split}
\psi(x,t_0+\Delta t)= & F^{-1}  [e^{-ik^2\Delta t} \\ 
& .F[ e^{i(2 | \psi_0 |^2 +  \beta^2 | \psi_0 |^4-2i\beta F^{-1}[ikF[| \psi_0 |^2]]  )\Delta t}\ \psi_0 ] ]
\end{split}
\label{eq13}
\end{equation}

Starting from the chaotic initial condition described above by Eq. (\ref{eq08}), the numerical solution of the KEE is obtained for later times by the SSFM. This form of the SSFM requires four fast Fourier transform (FFT) operations per time step. The number of spectral components are selected as $N=4096$ in order to make use of the FFTs efficiently. The time step is selected as low as $dt=10^{-4}$ which does not cause any stability problem in all runs.

\begin{figure}[ht!]
\begin{center}
   \includegraphics[width=3.4in]{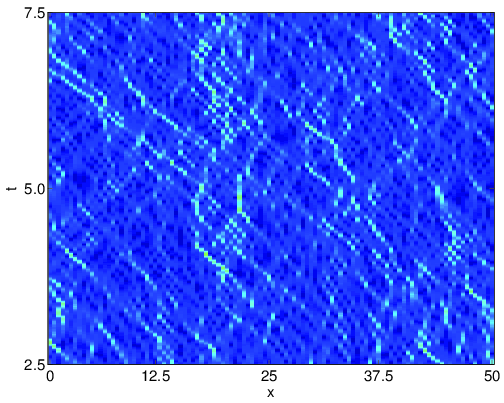}
  \end{center}
\caption{\small (Color online) A typical example of a chaotic wave field created by the KEE for $\beta=0.67$ starting with modulation instability. The filaments are propagating to left with average group velocity $v_{gr} \approx -6.5$.}
  \label{fig3}
\end{figure}

\begin{figure}[ht!]
\begin{center}
   \includegraphics[width=3.4in]{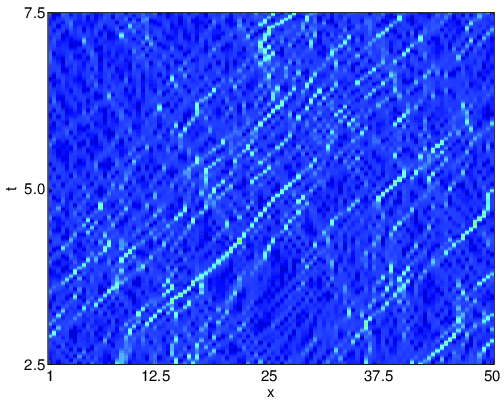}
  \end{center}
\caption{\small (Color online) A typical example of a chaotic wave field created by the KEE for $\beta=-0.67$ starting with modulation instability. The filaments are propagating to right with average group velocity $v_{gr} \approx 6.5$.}
  \label{fig4}
\end{figure}

Modulation instability started by the noise formulated above creates a chaotic wave field that starts from the initial plane-wave. An example of a wave field generated this way for $k=0.1$ is shown in Fig.\ref{fig3} for $\beta=0.67$ and in Fig. \ref{fig4} for $\beta=-0.67$. Checking the filaments in these figures, we can see that quintic and Raman-effect nonlinear terms produce an important skew angle relative to the ridge of the high waves in these chaotic fields. If $\beta > 0$ this skew angle is in the counter clockwise direction, else if $\beta < 0$ it is in the clockwise direction similar to the purely analytical results. 

The filaments are propagating to the right with average group velocity $ \left|v_{gr}\right| \approx 6.5$ in the Fig.\ref{fig4} however they are propagating to the left $\left| v_{gr} \right|  \approx 6.5$ in the Fig.\ref{fig3}. The different orientation for the skewed shape of the field is due to opposite signed $\beta$ values controlled by the Raman-effect term. Although the group velocity formula for the plane-wave can predict the magnitude of the velocity of the filaments, it deems insufficient to characterize propagation direction since the Raman-effect on the plane-wave dispersion relation drops out in its derivation. However one can realize that KEE is invariant under the  transformation $\beta \rightarrow -\beta $ and $x \rightarrow -x$, therefore filaments propagate in the opposite direction when the sign of the $\beta$ parameter changes. These properties of the KEE may be used to model and predict the skewed shape of the rogue waves encountered in practice.

The initial part of the field ($0< t <2.5$) is not shown in these figures as the deviations from the plane-wave solutions are very small and thus field amplitude is very close to 1 for all $x$ values. In contrast to the cubic NLSE case, the filaments of the KEE have a preferential direction of propagation with nonzero velocity, similar to other extension of the NLSE i.e. the Sasa-Satsumo equation \cite{Akhmediev2014}. This is due to the fact that the group velocity of the waves is not same as the phase velocity \cite{Akhmediev2014}.

\section{\label{sec:level1}Statistics of Big Waves}
The probability distribution of amplitudes ($\left| \psi \right|$) in the chaotic field provides important information about the wave field and about the rogue waves in particular. Therefore we obtain the probability density functions (PDFs) for various scenarios.
We numerically solve the KEE and simulate the chaotic field with in a spatial domain of $[-1000,1000]$. We discarded the initial modulation instability stages in our runs and used long spatial and temporal intervals to get statistical convergence.  We have divided the range of amplitudes, $\left| \psi \right|$, into $200$ bins in order to obtain relatively smooth curves and counted the number of maxima in each bin. Then by normalization we have obtained the PDFs. For each of the PDF plots, the data we have analyzed includes approximately one million maxima which allows us to obtain relatively smooth PDFs. 
\begin{figure}[htb!]
\begin{center}
   \includegraphics[width=3.4in]{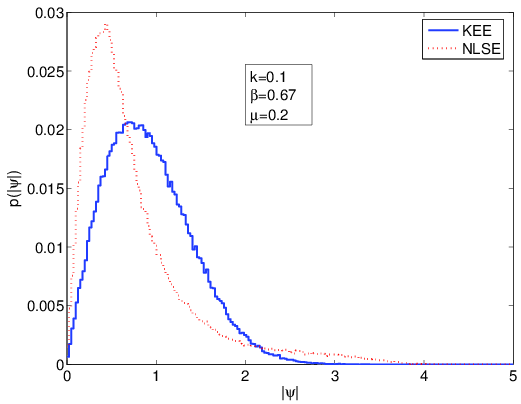}
  \end{center}
\caption{\small (Color online) PDF of the KEE ({\color{blue} ---})  vs. PDF of the cubic NLSE ({\color{red} - . -}) for $k=0.1$.}
  \label{fig5}
\end{figure}

\begin{figure}[htb!]
\begin{center}
   \includegraphics[width=3.4in]{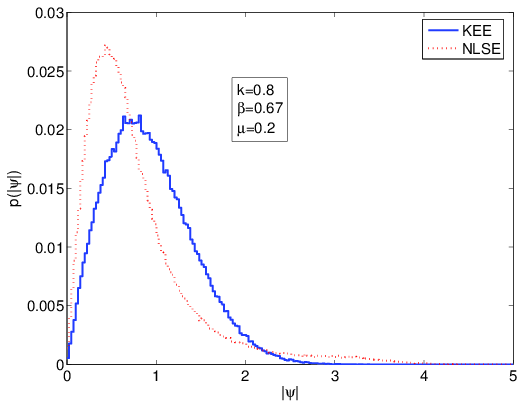}
  \end{center}
\caption{\small (Color online) PDF of the KEE ({\color{blue} ---})  vs. PDF of the cubic NLSE ({\color{red} - . -}) for $k=0.8$.}
  \label{fig6}
\end{figure}
\begin{figure}[htb!]
\begin{center}
   \includegraphics[width=3.4in]{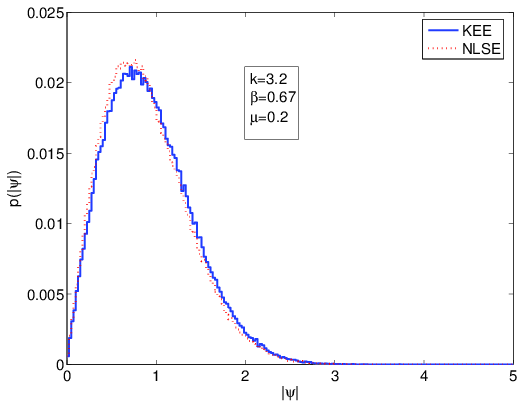}
  \end{center}
\caption{\small (Color online) PDF of the KEE ({\color{blue} ---}) vs. PDF of the cubic NLSE ({\color{red} - . -}) for $k=3.2$ .}
  \label{fig7}
\end{figure}

As Figs. \ref{fig5}-\ref{fig7} confirms, the PDFs for various values of the initial plane-wavenumber (k) show that smaller values of k result in higher probability of high-amplitude waves. Additionally, the comparison of PDFs of the KEE and NLSE fields show that for the smaller wavenumbers, more waves emerge in the KEE field than the NLSE field in the amplitude interval of approximately $[1,2.2]$. However the chaotic NLSE field produces significantly more waves in the amplitude interval of approximately $[2.2,5]$. Therefore it is possible to conclude that chaotic KEE field is less likely to produce rogue waves than the chaotic NLSE field. One possible explanation for this behavior is the contributions of the Raman-effect term. Although the KEE includes an additional quintic nonlinear term compared to the NLSE, the contribution of the Raman-effect term suppresses the contribution of quintic nonlinear term. Phase velocities of ultrashort waves are faster than longer waves, therefore the shorter waves do not contribute as a energy source for the rogue wave emergence in the field. However as the initial plane wavenumber gets bigger, there is no significant difference between the PDFs of the KEE and NLSE fields as Fig. \ref{fig7} confirms. These behaviors are quite similar to the results obtained for the Sasa-Satsumo equation in \cite{Akhmediev2014}.

\begin{figure}[htb!]
\begin{center}
   \includegraphics[width=3.4in]{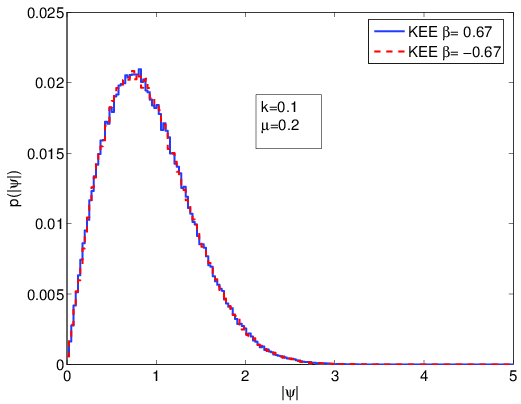}
  \end{center}
\caption{\small (Color online) PDF of the KEE for $\beta=0.67$ ({\color{blue} ---}) vs. PDF of the KEE for $\beta=-0.67$ ({\color{red} - . -}).}
  \label{fig8}
\end{figure}

We investigate the effect of the sign of the $\beta$ parameter on the probability of rogue wave formation in Fig.\ref{fig8}. It is clear that both the positive and negative $\beta$ values lead to similar PDFs thus it is possible to conclude that the sign of $\beta$ parameter controls only the skewed shape of wave field and do not alter the probability of rogue wave occurrence. 

\begin{figure}[htb!]
\begin{center}
   \includegraphics[width=3.4in]{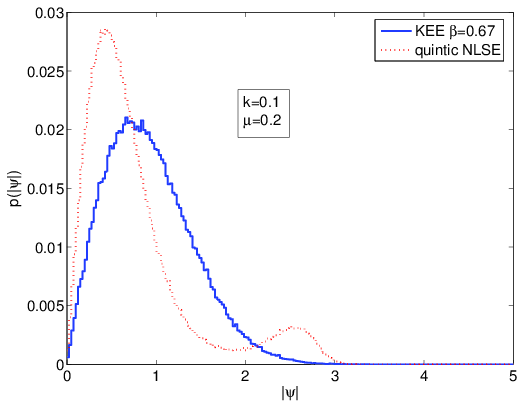}
  \end{center}
\caption{\small (Color online) PDF of the KEE ({\color{blue} ---}) vs. PDF of the quintic NLSE ({\color{red} - . -}).}
  \label{fig9}
\end{figure}

Next, we analyze the contribution of the Raman-effect term of the KEE on the probability of rogue wave formation. For this purpose we compare the PDF of the quintic NLSE which can be obtained by removing the Raman-effect term of the KEE and the PDF of the KEE in the Fig.\ref{fig9}. The quintic NLSE equation leads to a higher probability of amplitude occurrence in the interval of approximately $[2.2,5]$, which is significantly higher in the interval of approximately $[2.2,3.1]$. It is possible to conclude that the Raman-effect term lead to a lower probability of rogue wave formation. Underlying physical explanation is that this is caused by the self frequency shifts due to Raman-effect term and the faster propagation of ultrashort pulses than long pulses. This mechanism prohibits the energy acquisition of the rogue waves from the shorter pulses, therefore rogue waves are less likely to develop.

\section{\label{sec:level1}Conclusion}

In this paper, we studied the properties and statistics of the chaotic wave fields generated by the Kundu-Eckhaus equation. We have found that filaments of the chaotic fields generated by the modulation instability propagate with a velocity close to the magnitude of the average group velocity calculated from the dispersion relation for the plane-wave solution, however their propagation direction is controlled by the sign of the $\beta$ parameter. Our results can provide an explanation for the skewed shape of the rogue waves which may be observed in the practice. The calculation of the probability density functions for various values of the initial wavenumber showed that smaller values of k result in a higher probability of high amplitude waves. Our results have also demonstrated that the quintic nonlinear term in the evolution equation lead to higher probabilities of rogue wave occurrence in a chaotic wave field however the Raman-effect term in the Kundu-Eckhaus equation reduces the probability of rogue wave occurrence.

\end{document}